# A Extended Cosmological Cardy-Verlinde Formula

Thiago Gilberto do Prado

Marcos Cesar Verges

**Abstract**

E. Verlinde obtained a generalized formula for the entropy of a conformal field theory. For this we consider a (n+1) dimensional closed radiation dominated FLWR in the context of the holographic principle. In this work we construct a extension of the Cardy-Verlinde formula to positive cosmological constant spaces (dS spaces) with arbitrary topology

# Introduction

Bekenstein [Bekenstein] argued that for any system of energy $E$ and size $R$, the thermodynamic entropy is bounded by $S \leq S_B$, where $S_B = 2\pi ER$. However, Verlinde (Verlinde, 2000) proposed a normalization to the Bekenstein entropy bound. For this reason the Bekenstein entropy has a new form $S_B = \frac{2\pi ER}{n}$, where n is the spatial dimension of manifold. Also they propose a $n$ dimensional generalization of a (1+1) dimensional CFT (conformal field theory) entropy formula called the Cardy formula (Cardy, 1986), (Cardy, 1991). This generalization has the form

$$S = \frac{2\pi R}{n}\sqrt{E_c(2E - E_c)},$$

where $E_c$ is the Verlinde-Casimir energy and $E$ is the total energy. This entropy formula has a deep connection between the holographic principle (Bousso, 1999), (Bousso, 1999), (Hofft, 1993), entropy formulas for the CFT and the Friedman-Lemaitre-Robertson-Walker (FLRW) equations for a radiation dominated closed universe without cosmological constant. The FLRW equations without cosmological constant are given by

$$H^2 = \frac{16\pi G}{n(n-1)}\frac{E}{V} - \frac{1}{R^2},$$

$$\dot{H} = -\frac{8\pi G}{(n-1)}\left(\frac{E}{V} + p\right) + \frac{1}{R^2},$$

where $H = \frac{\dot{R}}{R}$ is the Hubble parameter, $\dot{H} = \partial_t H$, $G$ is the Newton constant and $p$ is the pressure. It was also found cosmological entropy written as

$$S_H = \frac{2\pi R}{n}\sqrt{E_{BH}(2E - E_{BH})},$$

where $S_H$ is the Hubble entropy bound and $E_{BH}$ is the Bekenstein-Hawking energy. This cosmological limit was found making use of FLRW equations without cosmological constant for a radiation dominated closed universe with entropy bounds and their valid conditions in phase transition between weak and strong self-gravity.

Some works, as (Rong, 2001), (Rong, 2002), (Setare 2004), (Setare, 2002),has tested the validity of the Cardy-Verlinde formula for different values of cosmological constants. For this, they have used black holes embedding in spacetime provided of cosmological constants, obtaining in general a favorable result to utilization of the Cardy-Verlinde formula to these cases. From these results we observed the necessity of the construction of a generalization of the Cardy-Verlinde formula valid to case of positive cosmological constant, arbitrary topology and for any energy condition.

In this work we intend to construct a extension of the Cardy-Verlinde formula to understand better its structure and behavior in the cases of positive cosmological constants, arbitrary topology of the universe and other conditions for the energy like weak and strong self-gravity.

# The Cosmological Entropy Bounds

From the knowledge that the black holes are the most entropic objects in the universe Bekenstein [Bekesntein,1981] propose a bound on the entropy of a macroscopic system. He argued that for a system with limited self-energy, the total entropy is less or equal than a multiple of the product of the energy and the linear size of the system. So the entropy bound is written as

$$S \leq S_B,$$

where

$$S_B = 2\pi ER. \qquad (1)$$

However, Verlinde [Verlinde] proposed that the Bekenstein entropy bound in the context of a closed radiation dominated FLRW universe with radius $R$ should be normalized. This new kind of Bekenstein entropy bound was written as

$$S_V = \frac{2\pi}{n} ER, \qquad (2)$$

where n is the spatial dimension of manifold. Here, for simplicity we will call the normalized Bekenstein entropy bound of Verlinde entropy bound. These entropy bounds are valid only in systems with limited self-gravity, where the following conditions are required

$$HR \leq 1, \quad E < E_{BH}.$$

Here $E_{BH}$ is the Bekenstein-Hawking energy.

For systems with strong self-gravity, the Bekenstein or Verlinde bounds are not valid. Hence for these systems are necessary a new entropy bound (in this system the possibility of black hole formation should be taken into account). This new entropy bound is called Bekenstein-Hawking entropy bound or holographic Bekenstein-Hawking entropy of a black hole with the size of the universe $(r_+ = R)$ and is given by

$$S_{BH} = (n-1)\frac{V}{4GR},$$

where $V$ is the volume of the universe. Thus when $S_B \leq S_{BH}$ and $HR \leq 1$, the self-gravity is weak, when $S_B \geq S_{BH}$ and $HR \geq 1$ the self-gravity is strong and exist a region of phase transition when $S_B = S_{BH}$ and $HR = 1$.

R. Brustein [Brustein, 1999] argued that the maximal entropy inside the universe is produced by black holes of the size of the Hubble horizon. Following holographic arguments he found that the total entropy should be less or equal than the Bekenstein-Hawking entropy of a Hubble size black hole. Therefore the Hubble entropy bound is written as

$$S_H = (n-1)\frac{HV}{4G}. \tag{3}$$

Where this entropy bound is valid only for $HR \geq 1$.

## Extended Cardy-Verlinde Formula (dS space)

In the (n+1) dimensional case, the FLRW equation with cosmological constant and arbitrary topology is given by

$$H^2 = \frac{16\pi G}{n(n-1)}\frac{E}{V} + \frac{\Lambda}{n^2(n-1)} - \frac{k}{R^2}, \tag{4}$$

where the cosmological constant in the n dimensional case (Witten, 1998) is

$$\Lambda = \frac{n(n-1)}{R^2}. \tag{5}$$

Now we will consider that the cosmological constant is positive, i.e., the dS case (Hawking and Ellis, 1973), (Wald, 1984) . Substituting the cosmological constant (5) in (4) we find the expression of the FLRW equation to the dS case

$$H^2 = \frac{16\pi G}{n(n-1)}\frac{E}{V} + \frac{1}{nR^2} - \frac{k}{R^2}.$$

In our deduction we will be interested not only in the case of weak ( $HR \leq 1$ ) or strong ( $HR \geq 1$ ) self-gravity or phase transition ( $HR = 1$ ) (Verlinde, 2000), but in all cases. So we will define $HR = \sqrt{c}$ (Nojiri et al, 2001), where $c$ is a constant of order one. When we write this, the total energy $E$ becomes equal to the Bekenstein-Hawking energy $E_{BH}$ in the phase transition. We have

$$(HR)^2 = \frac{16\pi G}{n(n-1)} \frac{E_{BH} R^2}{V} + \frac{1}{n} - k.$$

After some algebra over the before expression we find that the Bekenstein-Hawking energy is expressed as

$$E_{BH} = \frac{[cn - 1 + nk](n-1)V}{16\pi G R^2}. \tag{6}$$

We can see that the Bekenstein-Hawking energy is proportional to R$^{-2}$ in agreement with others references (Nojiri et al, 2001), (Wang et al, 2001).

Now we have two different entropy bounds: the Bekenstein entropy bound $S_B = 2\pi ER$ and the Verlinde entropy bound $S_V = \frac{2\pi}{n} ER$. The only difference between this entropies is the normalization factor that depends of the spatial dimension. So we want to know if exist some difference among the Cardy-Verlinde formula obtained from the Bekenstein entropy bound and the other obtained from the Verlinde entropy bound. We will test these two entropy bounds in this text.

It Is known that in the case of phase transition these entropy bounds are equal to the Bekenstein-Hawking entropy bound $S_{BH}$ (Nojiri et al, 2001). This fact will be used several times in this text.

The first test: Bekenstein entropy bound $S_B = 2\pi ER$. With this condition we find

$$S_{BH}^B = \frac{[cn - 1 + nk](n-1)V}{8GR}, \tag{7}$$

The proportionality of $S_{BH}^B$ with $\frac{V}{R}$ is in agreement with references (Wang et al, 2001), (Nojiri et al, 2001). Making use of (3) ,(1) and (7) in the FLRW (4) we can rewrite the FLRW equation as

$$S_{HB}^2 = \frac{4}{n[cn-1+nk]} \left( S_B S_{BH}^B - \frac{(S_{BH}^B)^2 (nk-1)}{[cn-1+nk]} \right) \tag{8}$$

If we write (8) in terms of the energy, we find

$$S_{HB} = 2\pi R \sqrt{\frac{4E_{BH}}{n[cn-1+nk]}\left(E - \frac{E_{BH}(nk-1)}{[cn-1+nk]}\right)}, \tag{9}$$

that is the general cosmological Cardy-Verlinde formula for any value of $c$, $n$ and $k$.

The second test: Verlinde entropy bound $S_V = \frac{2\pi}{n} ER$. Using the normalized Bekenstein condition, we found

$$S_{BH}^V = \frac{[cn-1+nk](n-1)V}{8nGR}. \tag{10}$$

The only difference between (7) and (10) is the normalization factor $n$. Using (3), (2) and (10) in the FLRW (4) we can (like in the first test) rewrite the FLRW equation as

$$S_{HV}^2 = \frac{4n}{[cn-1+nk]}\left(S_B S_{BH}^V - \frac{(S_{BH}^V)^2(nk-1)}{[cn-1+nk]}\right). \tag{11}$$

Now we note that the difference between (8) and (11) is the normalization factor n again. Just like before we can write (11) in terms of the energy,

$$S_{HV} = 2\pi R \sqrt{\frac{4E_{BH}}{n[cn-1+nk]}\left(E - \frac{E_{BH}(nk-1)}{[cn-1+nk]}\right)}, \tag{12}$$

that is the general cosmological Cardy-Verlinde formula for any $c$, $n$ and $k$ obtained using the Verlinde entropy bound. We can verify that both expressions, the Cardy-Verlinde holographic entropy in the cosmological case (9) and (12), obtained using the Bekenstein entropy bound and the Verlinde entropy bound are exactly the same. So we verify that the normalization factor $n$ of the Verlinde entropy bound is absorbed. Therefore there are no difference between Bekenstein entropy bound or Verlinde entropy bound if our objective is found an entropy formula for a CFT in (n+1) dimensional case.

## Conclusions

We found that there are no difference between the use of the Bekenstein entropy bound in its original form or the use of this entropy bound in its normalized form to construct the cosmological Cardy-Verlinde formula. Hence the normalization factor for the Bekenstein entropy bound proposed by Verlinde in (Verlinde, 2000) is not necessary. So the Bekenstein entropy bound is always valid for a Conformal Field Theory in any dimension without modification.

We can verify for direct substitution that the extension of Cardy-Verlinde formula reduces to the original Cardy-Verlinde formula when $\Lambda = 0$, $k = 1$ and, $c = 1$ that is exactly the conditions used by Verlinde in (Verlinde, 2000) for the attainment of (n+1) dimensional generalization of Cardy formula. This simply verification shows that the extension is correct, but a deeper analysis of this formula is necessary to compare the possible corrections in problems involving a positive cosmological as the entropy of dS Black holes that had been solved using the usual Cardy-Verlinde formula.